\documentclass[pra,showpacs,aps,amssymb,amsmath,nofootinbib,notitlepage,superscriptaddress,twocolumn,longbibliography]{revtex4-2}

\usepackage{graphicx,graphics,epsfig,times,bm,mathrsfs}
\usepackage[normalem]{ulem}
\usepackage{subcaption}
\usepackage{xfrac}
\usepackage{mathtools}
\usepackage{amsmath,amssymb,amsfonts}
\usepackage{varwidth}
\usepackage{babel}
\usepackage{hhline}
\usepackage{multirow}
\usepackage{pgf}
\usepackage{physics}
\usepackage{bbm}
\usepackage{float}
\usepackage{microtype}
\usepackage{booktabs}
\usepackage[T1]{fontenc}
\usepackage[utf8]{inputenc}
\usepackage{algorithm}
\usepackage[noend]{algpseudocode}
\usepackage{cancel}
\usepackage[hidelinks]{hyperref}

\usepackage{lineno}

\begin{document}

\title{Proof-of-principle long-distance Sagnac twin-field quantum key distribution network}

\author{Reem Mandil}
\email{reem.mandil@mail.utoronto.ca} 
\affiliation{Centre for Quantum Information and Quantum Control, Department of Physics, University of Toronto, Toronto, Ontario, M5S 1A7, Canada}

\author{Yen-An Shih}
\affiliation{Centre for Quantum Information and Quantum Control, Department of Electrical and Computer Engineering, University of Toronto, Toronto, Ontario, M5S 3G4, Canada}
\affiliation{Department of Physics and Center for Quantum Science and Technology, National Tsing Hua University, Hsinchu 300, Taiwan} 

\author{Abhay Verma}
\affiliation{Centre for Quantum Information and Quantum Control, Department of Electrical and Computer Engineering, University of Toronto, Toronto, Ontario, M5S 3G4, Canada}

\author{Li Qian}
\affiliation{Centre for Quantum Information and Quantum Control, Department of Electrical and Computer Engineering, University of Toronto, Toronto, Ontario, M5S 3G4, Canada}

\author{Hoi-Kwong Lo}
\affiliation{Centre for Quantum Information and Quantum Control, Department of Electrical and Computer Engineering, University of Toronto, Toronto, Ontario, M5S 3G4, Canada}
\affiliation{Department of Physics, National University of Singapore, 2 Science Drive 3, Singapore 117551}
\affiliation{Centre for Quantum Technologies, National University of Singapore, 21 Lower Kent Ridge Road, Singapore 119077}
\affiliation{Quantum Bridge Technologies, Inc., 100 College Street, Toronto, Ontario, M5G 1L5, Canada}

\date{\today}

\begin{abstract}
Twin-field (TF) quantum key distribution (QKD) offers a promising approach to long-distance QKD networks due to its superior performance over large channel losses. Due to specialized hardware requirements, nearly all long-distance TFQKD demonstrations have only two users exchanging keys, rather than a network with three or more users. In this work, we experimentally demonstrate a proof-of-principle three-user-pair Sagnac TFQKD network spanning 127-km using single-photon avalanche detectors without any active phase stabilization or postcompensation. We implement efficient procedures for maintaining polarization stability and circumventing Rayleigh backscattering noise to achieve a stable Sagnac interference visibility of $93\pm1$\% over one hour. Using measured data for the signal basis and simulated data for the decoy basis, an asymptotic secure key rate of $1.398\times10^{-5}$~bits per pulse is achieved over an asymmetric communication channel with 102-km fiber and 45-dB overall loss. To our knowledge, this is the first TFQKD network without active phase stabilization or postcompensation achieved over long fibers. Our results represent a highly practical and cost-effective approach to long-distance QKD networks.
\end{abstract}

\maketitle

\section{Introduction}\label{sec:intro}
Quantum key distribution (QKD) offers information-theoretic security by utilizing quantum carriers that cannot be copied or intercepted without alerting the legitimate parties~\cite{bb84,e91,Xu2020,Portmann2022}. While initially established between just two remote users, QKD needs to expand into robust, multi-user networks~\cite{Townsend1997,Elliot2005,Peev2009,Sasaki2011,Frohlich2013,Tang2016,Chen2021,Mandil2023} to incentivize broad commercial adoption. The challenges faced by such networks are twofold: the ability to add/drop users in a network easily and seamlessly, and the transmission of fragile quantum signals over long distances of optical fiber and lossy network switching/connection points.

Twin-field QKD (TFQKD)~\cite{Lucamarini2018}, the most recent variant of measurement-device-independent (MDI) QKD~\cite{Lo2012mdi}, is best positioned to overcome the loss challenge due to its $O(\sqrt{\eta})$ scaling with channel transmittance $\eta$. Furthermore, due to its MDI architecture, fully connecting $N$ users in a TFQKD network requires only $O(N)$ fiber links and a single pair of detectors, providing low-cost scalability as compared to a standard point-to-point QKD network which requires $O(N^2)$ fiber links and pairs of detectors. From this viewpoint, TFQKD is particularly suitable for large network implementation where quantum signals are faced with loss not only due to fiber attenuation, but also due to lossy network switches and performance monitoring devices. However, while tremendous progress has been made to extend TFQKD distance, e.g., over 1,000-km fiber~\cite{Liu2023}, there remain significant hurdles to developing a viable, practical, and low-cost TFQKD \textit{network}.

Park \textit{et al.}~\cite{Park2022} demonstrated a TFQKD network using superconducting nanowire single-photon detectors (SNSPDs) in a Sagnac-based plug-and-play (PnP) architecture with 160-km fiber. Common-path stable interference was not observed in their experiment and therefore they used a phase postcompensation method to compensate for the fiber phase drift. Another work, by Clason \textit{et al.}~\cite{Clason2026}, uses a similar Sagnac PnP scheme for passive stabilization and demonstrates a two-user TFQKD system over 110-km of installed fiber using SNSPDs. Recently, Zheng \textit{et al.}~\cite{Zheng2026} demonstrated a TFQKD network using SNSPDs where Charlie and the user stations are implemented using integrated photonics. Charlie generates frequency combs with Hz-level linewidths that are distributed to all users to phase-lock their sources. Note that Ref.~\cite{Zheng2026} essentially demonstrates a network with two central stations, each connected to multiple users that are separated by negligible distances. While they are one type of multiuser network, they are not true multi-hop user networks. 

In this paper, we demonstrate a fiber-based TFQKD multi-hop ring network where three users are separated by long fibers with unequal lengths, as would be the case in a real-world network. Furthermore, the entire network operates with inexpensive single-photon avalanche detectors (SPADs) and a single commercial distributed feedback laser, without the need for frequency locking thanks to Sagnac interference. This work differs from our previous TFQKD network demonstration~\cite{Zhong2022PRApp} in that long fibers, not variable optical attenuators (VOAs), are used between users, resulting in two significant challenges: 1) severe Rayleigh backscattering obscuring quantum signals, which travel bidirectionally in a ring network; and 2) higher phase and polarization instability caused by long fibers. Built upon a feasibility study~\cite{Mandil2025} demonstrating high visibility interference of quantum signals in a long fiber Sagnac loop, this work demonstrates quantum key generation between pair-wise users in a three-user network spanning 127-km. As a proof-of-principle experiment, quantum key is experimentally generated in the key-generation basis (which is critically dependent on interference visibility), while the decoy state transmission (insensitive to visibility) is simulated with real channel losses. We obtain positive secure key rates for both pairs, including in the finite data-size regime for one pair. To the best of our knowledge, our results represent the first TFQKD network without active phase stabilization or postcompensation achieved over long fibers.

\section{Protocol}\label{sec:protocol} 
In this work, we adopt the finite-size security analysis in Ref.~\cite{Wang2020} to optimize the key rate for each user pair. This security analysis extends the TFQKD protocol in Ref.~\cite{Curty2019} to the scenario of asymmetric channels between Charlie and each user and considers a security failure probability of $10^{-7}$. The protocol for TFQKD with asymmetric channels has been previously demonstrated in Refs.~\cite{Zhong2021,Zhong2022PRApp} and is comprised of the following five steps. (1) Two users independently select a basis, choosing the $X$ (signal) basis with probability $P_X$ and the $Z$ (decoy) basis with probability $P_Z=1-P_X$. If the $X$ basis is chosen, the user prepares a WCP with a preselected global phase and intensity $s$ according to probability $P_s=P_X$. The user selects between a bit value of $0$ or $1$ at random, to add a $0$ or $\pi$ phase, respectively, to their WCP. If the $Z$ basis is chosen, the user prepares a phase-randomized WCP with intensity chosen from a set of decoy settings $\{\mu,\nu,\omega\}$, where $\omega$ is the vacuum state, according to probabilities $P_\mu, P_\nu, P_\omega$. (2) The two users send their prepared WCPs to an untrusted central relay, Charlie. (3) At Charlie’s station, the incoming WCPs interfere with each other at a 50:50 beam splitter followed by two single-photon detectors, $D_0$ and $D_1$. A measurement event is considered successful if one and only one detector clicks. (4) Charlie announces the results of the successful measurement events (i.e., which detector clicked) and the two users declare which basis they selected. (5) Based on the information announced, the two users distill the secret key. The instances where both users chose the $X$ basis are used to form the raw key and determine the quantum bit error rate (QBER). The instances where both users chose the $Z$ basis are used to implement the decoy-state method and estimate the phase error rate. The key rate is calculated from the observed gains (i.e., detection rates) in the $X$ and $Z$ bases, the observed QBER in the $X$ basis, and the estimated phase error rate.

Note that, in the first step, the signal intensity may be different for each user~\cite{Wang2020}. This strategy compensates for the channel-loss asymmetry and achieves higher key rates than fixing the signal intensity to be the same for each user or padding loss to make channel losses symmetric~\cite{Zhong2021}. The decoy intensities are the same for both users since the decoy-state method and phase error rate estimation are based on photon-number yields in the $Z$ basis and are unaffected by an asymmetry of intensities arriving at Charlie~\cite{Wang2020}.

We note that in this protocol, it is only the signal basis that relies on good interference visibility. Therefore, as a proof-of-principle demonstration, we implement only the signal basis in experiment and use a simulated channel model to obtain the gains in the decoy basis. In this work, we focus on showing that even over long fibers, we can achieve sufficiently low bit error rates for key generation (including in the asymmetric channel and finite-size regimes) with remarkable simplicity and cost-effectiveness. Our proof-of-principle implementation is adequate for this purpose.

\section{Experiment}\label{sec:experiment}
\begin{figure*} [htbp]
    \begin{center}
       \includegraphics[width=0.95\textwidth]{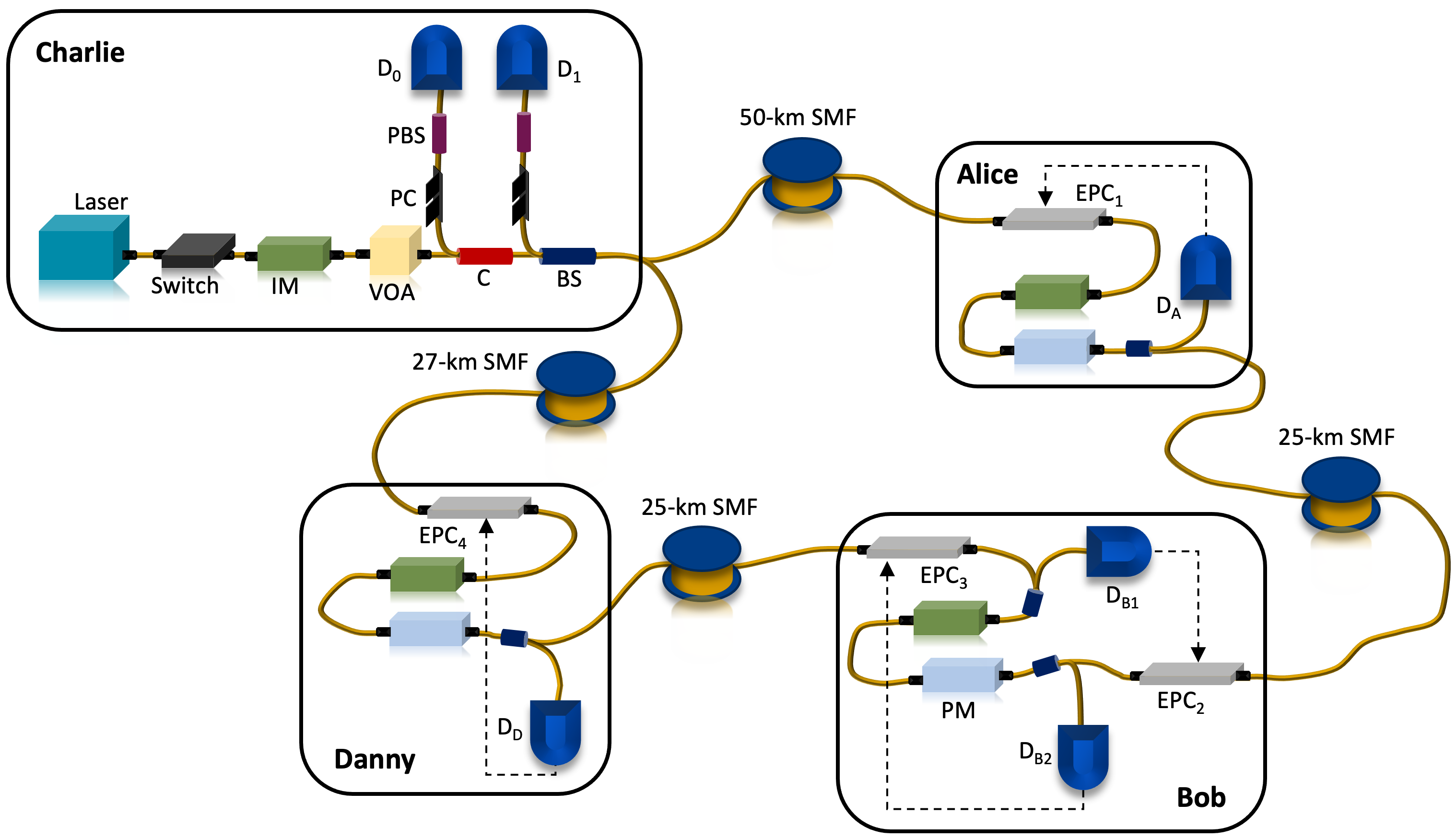} 
    \end{center}
    \caption{Experimental setup of twin-field quantum key distribution network based on a Sagnac interferometer. Charlie passes light from a continuous-wave laser to a $1\times1$ optical switch to generate an on-off burst pattern. An intensity modulator (IM) and variable optical attenuator (VOA) are used to generate pulses that are launched into a fiber ring via a 50:50 beam splitter (BS). Three users (Alice, Bob, and Danny) are connected in the network. Any two users may perform key generation while the third user remains inactive. The active users employ an IM and a phase modulator (PM) to set the intensity and phase of their designated pulses. Interference of the counterpropagating paths is measured by single-photon avalanche detectors (SPADs) $D_0$ and $D_1$. Polarizing beam splitters (PBSs) are used to filter approximately half of the backscattering noise from the signal. The fiber segments between the active users and between Charlie and each user are polarization-aligned using electronic polarization controllers (EPCs). A 99:1, 95:5, 99:1, and 95:5 BS is used to direct a small portion of the light to SPADs $D_A$, $D_{B1}$, $D_{B2}$, and $D_D$, respectively. Feedback from these intensity monitors is used to drive the EPCs. C: circulator; PC: polarization controller.} \label{fig:setup}
\end{figure*}

Our Sagnac-based TFQKD network is depicted in Fig.~\ref{fig:setup}. Light from a cw distributed feedback laser diode (1545.3 nm) passes through a polarizer (not pictured) followed by a $1\times1$ optical switch (extinction ratio $>$~25 dB) driven by a square rf signal that determines when it is open or closed, thereby generating a burst pattern. Burst-patterning is necessary to achieve high signal-to-noise ratio (SNR) in the presence of Rayleigh backscattering noise~\cite{Mandil2025, Xingye2023}. After the switch, an intensity modulator (IM) is used to generate pulses at a repetition rate of 10~MHz with 900-ps pulse width. A VOA is used to adjust the source intensity before pulses are launched into a fiber ring via a symmetric beam splitter. Inside the Sagnac loop, there are three users: Alice, Bob, and Danny. Each user's station contains an IM and a phase modulator (PM) for encoding, and EPCs that are used to maintain polarization alignment inside the fiber ring. A 99:1, 95:5, 99:1, and 95:5 beam splitter is used to direct a small portion of the light to free-run SPADs (ID220 or IDQube) $D_A$, $D_{B1}$, $D_{B2}$, and $D_D$, respectively. Details on our polarization stabilization scheme are presented in Sec.~\ref{sec:polarization}.

Similar to our previous work in Ref.~\cite{Zhong2022PRApp}, any two users may perform key generation while the third user remains inactive. In the Sagnac loop, pulses travel along either a clockwise or counterclockwise path. In a given key generation session, each active user modulates only one path, namely the one that has already traversed the other active user. Pulses on the other path pass through without modulation. For instance, in the Alice-Bob pair, Alice will modulate the counterclockwise traveling pulses while Bob modulates the clockwise traveling pulses. The fiber length at each user station can be adjusted such that counterpropagating pulses never overlap at any modulator~\cite{Zhong2022PRApp}. After traversing the full loop, the counterpropagating paths interfere at Charlie's beam splitter. The output of the interference is recorded by two free-run SPADs (IDQube), $D_0$ and $D_1$, with an efficiency of 10\% and a dark count rate of about $3\times10^{-7}$ per pulse. Polarizing beam splitters are used to filter approximately half of the backscattering noise from the signal, since the polarization of the scattered light is random. Measurements are recorded using a time-interval analyzer (HydraHarp400). A high-speed multichannel arbitrary waveform generator (Keysight M8195A) provides the synchronized rf signals to the time-interval analyzer and to each of the modulators. The delay time of the electrical signal to each modulator is calibrated such that the modulation windows overlap with the designated pulses. 

Unlike in our previous demonstration where the total length of the Sagnac loop was 10-km~\cite{Zhong2022PRApp}, here the total length is 127-km, made up of four spools of SMF-28 ultra-low-loss (ULL) fiber. Between Alice and Charlie, there is 50-km; between Danny and Charlie there is 27-km. The distance between Bob and Charlie depends on who Bob is communicating with. In the Alice-Bob pair, there is 52-km between Bob and Charlie; In the Danny-Bob pair, there is 75-km. 

\section{Polarization Stabilization}\label{sec:polarization}
\begin{figure*} [htbp]
    \begin{center}
       \includegraphics[width=\textwidth]{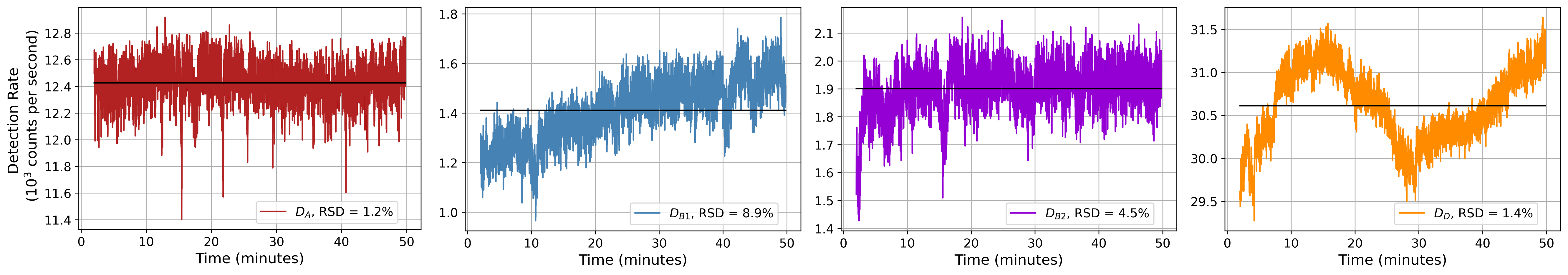} 
    \end{center}
    \caption{Counts registered at single-photon detectors $D_A$, $D_{B1}$, $D_{B2}$, and $D_D$ during one hour of active polarization control. $D_A$ and $D_{B1}$ are used to monitor the clockwise traveling light. $D_D$ and $D_{B2}$ are used to monitor the counterclockwise traveling light. The mean (solid line) and standard deviation are given in counts per second by $D_A$: $12430\pm150$; $D_{B1}$: $1410\pm120$; $D_{B2}$: $1900\pm90$; $D_D$: $30610\pm430$.} \label{fig:polmonitors}
\end{figure*}

In our experiment, the interfering light fields should have identical polarizations in order to maximize the interference visibility. Furthermore, for accurate encoding, the polarizations of the counterpropagating light fields should also be aligned to the slow axis of the lithium niobate crystal inside the modulators. Disturbances in the surrounding environment that cause the fiber birefringence to fluctuate~\cite{Barlow1983} will result in polarization misalignment between the interfering fields because the fiber birefringence differently changes the polarization of the clockwise and counterclockwise traveling light~\cite{Mecozzi2011}. Therefore, active polarization control is needed to maintain long-term interferometric stability. 

To achieve this goal, each user employs a PM fitted with a polarizer and an IM with a polarization-maintaining fiber input. This configuration guarantees proper alignment inside the modulators. Then, the fiber segments between the users and between Charlie and each user are polarization-aligned to ensure maximum interference visibility. More precisely, the intensity of light passing through the user station is monitored using a fiber tap and a SPAD. This intensity signal is used to drive an EPC that will adjust its setting to increase the intensity. 

In Fig.~\ref{fig:setup}, Alice monitors the intensity of clockwise traveling light passing through her station with detector $D_A$ which is used to drive the EPC in the fiber segment between Charlie and Alice (EPC$_1$). Bob monitors the intensity of clockwise traveling light passing through his station with detector $D_{B1}$ which is used to drive the EPC in the fiber segment between Alice and Bob (EPC$_2$). Similarly, Bob monitors the intensity of counterclockwise traveling light passing through his station with detector $D_{B2}$ which is used to drive the EPC in the fiber segment between Danny and Bob (EPC$_3$). Danny monitors the intensity of counterclockwise traveling light passing through his station with detector $D_{D}$ which is used to drive the EPC in the fiber segment between Charlie and Danny (EPC$_4$). Note that only one of the users needs to monitor both directions. This feature arises by invoking that, in a given fiber segment, the polarizations of the counterpropagating fields undergo unitary transformations that are transpose to one another~\cite{Mecozzi2011}. The theory and algorithm behind our polarization stabilization scheme is detailed in Appendix~\ref{app:pol_theory}. 

Fig.~\ref{fig:polmonitors} shows the intensity at monitors $D_A$, $D_{B1}$, $D_{B2}$, and $D_D$ during one hour of active polarization control. The intensity is stable with a relative standard deviation (standard deviation divided by the mean, calculated over the entire time window) of 1.2\%, 8.9\%, 4.5\%, and 1.4\% at $D_A$, $D_{B1}$, $D_{B2}$, and $D_D$, respectively. The small variations in the intensity are predominantly due to random noise. Occasional dips are observed, which may be due to environmental disturbance near the fiber. Small systematic drifts are visible at $D_D$, which are likely artifacts of our polarization stabilization algorithm. The intensity at $D_{B1}$ shows a steady increase for the first twenty-five minutes, likely due to longer fiber preceding this monitor.

\section{Results}\label{sec:results}
\begin{figure} [b]
    \begin{center}
       \includegraphics[width=0.99\linewidth]{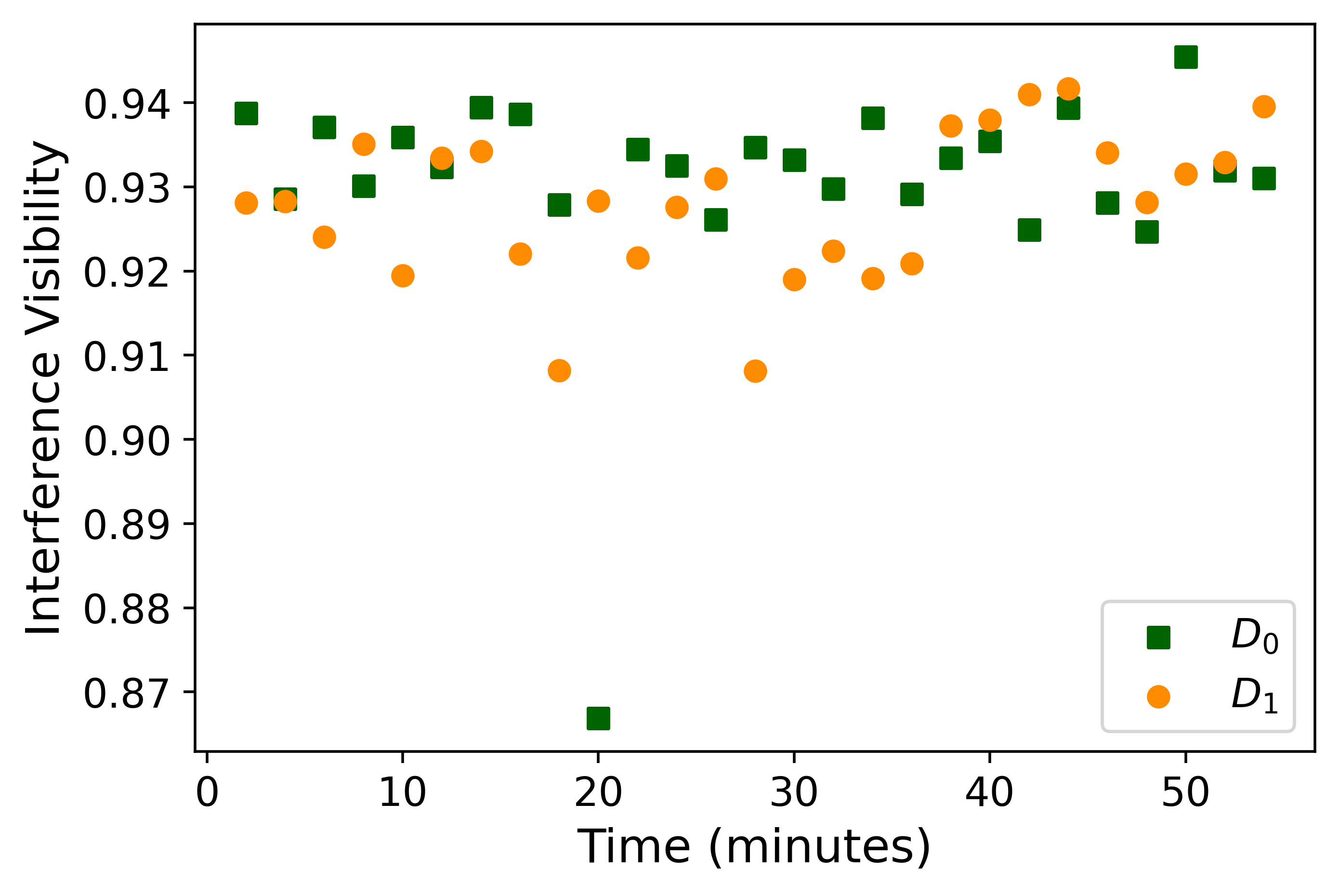} 
    \end{center}
    \caption{Sagnac interference visibility in detectors $D_0$ and $D_1$ over one hour with active polarization control.} \label{fig:visvtime}
\end{figure}

To demonstrate the long-term interferometric stability of our Sagnac-based TFQKD network, we record the interference visibility over one hour as shown in Fig.~\ref{fig:visvtime}. Measurements are recorded using an integration time of 60-s. Between each measurement, we use Alice's PM to alternate which detector ($D_0$ or $D_1$) receives total constructive interference. Then, the visibility in each detector is calculated by \begin{equation}\label{eq:visibility}
    V = \frac{I_{\text{max}}-I_{\text{min}}}{I_{\text{max}}+I_{\text{min}}},
\end{equation} where $I_{max}$ is the pulse intensity during constructive interference and $I_{min}$ is the pulse intensity during destructive interference. We achieve a steady average visibility of $93\pm1$\% over one hour. With the exception of one aberrant data point at the twenty minute mark where the visibility drifts to 86.7\% in $D_0$, the visibility in both detectors stays between 90.8\% and 94.5\% with no observable downward trend. We note that strong pulses are used for the interference visibility measurements in order to assess system stability when the background noise has a small impact on the SNR.

Table~\ref{tab:tfqkd} shows the results of our TFQKD demonstration for two user pairs in the network. The signal intensity ($s$) is different for each user while the decoy intensities ($\mu, \nu$) and probabilities of sending signal and decoy states ($P_s, P_{\mu}, P_{\nu}$) are the same. The probability $P_{\omega}$ is given by $1-P_s-P_{\mu}-P_{\nu}$. Since we implement only the signal basis in experiment (as described in Sec.~\ref{sec:protocol}), only the signal intensities and gains are measured parameters, while the decoy intensities and gains are those obtained by a key rate optimization for the given channel~\cite{Wang2020}. The sending probabilities are also optimized parameters. The QBER is given for detectors $D_0$ and $D_1$. Key rates in the infinite-data and finite-data regimes are shown along with the repeaterless bound~\cite{Pirandola2017} for a given channel loss. The detector efficiency (10\%) is attributed as part of the loss between Charlie and each user since our low-efficiency detector can be modeled as a loss element followed by a unity-efficiency detector. 

\begin{table*}[htbp]
  \centering
  \caption{Parameters and key rates (bits per pulse) in three-user-pair Sagnac TFQKD network. The loss and distance of the channel between Charlie and each user are listed. The detector efficiency (10\%) is attributed as part of the loss between Charlie and each user. Here $s_1,s_2$ are signal intensities (average photon number per pulse); $\mu,\nu$ are decoy intensities; $P_s,P_\mu,P_\nu$ are probabilities of sending intensities; QBER is quantum bit error rate. The data size for the finite-data case is $10^{10}$.}
  \label{tab:tfqkd}
  \resizebox{\textwidth}{!}{
  \begin{tabular}
  {ccccccccccccccc}
  \toprule
  \textbf{Users}&\multicolumn{2}{c}{\textbf{Channel (User 1+User 2)}}&\multicolumn{7}{c}{\textbf{Intensity}}&\multicolumn{2}{c}{\textbf{QBER}}&\multicolumn{3}{c}{\textbf{Key rate} ($\times10^{-5}$ bits per pulse)}\\
  \cmidrule(lr){2-3}\cmidrule(lr){4-10}\cmidrule(lr){11-12}\cmidrule(lr){13-15}
  & Loss (dB) & Distance (km) & $s_1$ & $s_2$ & $\mu$ & $\nu$ & $P_s$ & $P_\mu$ & $P_\nu$ & $D_0$ & $D_1$ & Infinite data & Finite data & Repeaterless bound \\
  \midrule
  Alice+Bob & 18.6+26.1 & 50+52 & 0.0049(1) & 0.0226(2) & 0.4904 & 0.1307 & 0.7480 & 0.0253 & 0.0653 & 7.50\% & 5.49\% & 1.398 & 0 & 4.889 \\
  Alice+Danny & 18.6+15.5 & 50+27 & 0.0223(1) & 0.0116(1) & 0.4752 & 0.1360 & 0.8600 & 0.0160 & 0.0432 & 6.02\% & 4.03\% & 7.190 & 0.4408 & 56.14 \\
  \bottomrule
  \end{tabular}
  }
\end{table*}

For the Alice-Bob pair, which consists of a 102-km channel with 45-dB overall loss, we achieve a QBER (average of $D_0$ and $D_1$) of 6.49\%. Detector $D_0$ has a higher QBER since pulses undergo roughly 1-dB more loss due to the circulator. We obtain a key rate in the infinite-data case of $1.398\times10^{-5}$~bits per pulse. For the Alice-Danny pair, which consists of a 77-km channel with 34-dB overall loss, we achieve a QBER (average of $D_0$ and $D_1$) of 5.02\%. We obtain a key rate of $7.190\times10^{-5}$~bits per pulse in the infinite-data case and a key rate of $4.408\times10^{-6}$~bits per pulse in the finite-date case.
Currently, the SNR in the case of the Bob-Danny pair is insufficient for secure key generation. The overall loss for this pair is roughly 49-dB. The additional loss as compared to the Alice-Bob pair is mainly due to the difference in loss between Danny’s station (low-loss modulators) and Alice’s station (modulators have higher loss). With this roughly 4-dB extra loss, and with the current background noise level in our system, we expect the QBER to exceed 10\%. Strategies for recovering SNR in our system are discussed in Sec.~\ref{sec:discussion}. Prior to this network demonstration, we tested our long-fiber system under a symmetric configuration with only two users (Alice and Bob) in the loop. The results are found in Appendix~\ref{app:twouser}. 

\section{Discussion}\label{sec:discussion}
While our system with intrinsic phase stability is unlikely to achieve transmission distances as long as other setups for TFQKD, it has the unique advantages of being low-cost and scalable to many users, representing a significant leap towards the practical implementation of metropolitan quantum networks. Here, we provide a detailed comparison between Sagnac TFQKD and other TFQKD systems with regards to ease of implementation and security.

Neglecting birefringence effects (which are addressed through polarization compensation), the phase difference $\Phi(t)$ of the two light fields at Charlie evolves as~\cite{Li2023}    
\begin{equation}\label{eq:phase_diff}
    \Phi(t) = \phi_0 + 2\pi\nu_0t + \Delta\phi_{laser}(t) + \Delta\phi_{fiber}(t),
\end{equation} where $\phi_0$ is the difference in the initial phases and $\nu_0$ is the initial frequency difference between Alice's and Bob's lasers (`beat-note frequency'). The terms $\Delta\phi_{laser}(t)$ and $\Delta\phi_{fiber}(t)$ are the phase fluctuations (random noise) introduced during channel transmission by laser phase noise and fiber phase noise, respectively. The term $\Delta\phi_{laser}(t)$ is caused by laser frequency drift and mode-hopping; $\Delta\phi_{fiber}(t)$ is caused by physical length and refractive index fluctuations in fiber.

Between the laser phase noise and the fiber phase noise, the former is more problematic when Alice and Bob employ conventional laser sources (several kHz in linewidth). In the best-case scenario where the two laser sources are frequency-locked (meaning their phase noise replicates that of a shared reference source), $\Delta\phi_{laser}(t)$ is proportional to $(L_B-L_A)\Delta v(t)$, where $L_B$ and $L_A$ are the channel lengths between Charlie and each user and $\Delta v(t)$ is the frequency fluctuation (which generally correlates with linewidth). Therefore, even a small $\Delta v(t)$ can lead to a significant phase fluctuation if the channel lengths are asymmetric. Without laser-locking, $\Delta\phi_{laser}(t)$ can approach $(L_B+L_A)\Delta v(t)$ in the worst case, meaning a small $\Delta v(t)$ produces a large phase fluctuation even if the channels are symmetric. Therefore, as the channel length or laser linewidth increases, the laser phase noise will dominate the phase fluctuation. 

In order to demonstrate secure key generation beyond the repeaterless bound, most of the research effort in TFQKD experiments~\cite{Zhong2019, Zhong2021, Zhong2022PRApp, Minder2019, Wang2019, Liu2019, Fang2020, Chen2020, Chen2021twin, Liu2021, Pittaluga2021, Wang2022, Park2022, Liu2023, Li2023, Zhou2023, Chen2024, Pittaluga2025, Zheng2026} is aimed at tackling the phase difference (Eq.~\ref{eq:phase_diff}) over hundreds of kilometers of fiber. The experiments in Refs.~\cite{Minder2019, Wang2019, Liu2019, Fang2020, Chen2020, Chen2021twin, Liu2021, Pittaluga2021, Wang2022, Liu2023, Li2023, Zhou2023, Chen2024, Pittaluga2025, Zheng2026} mitigate each of the laser and fiber phase noise either in real time or through postprocessing techniques. Laser phase noise is reduced in real time through laser-locking methods such as optical phase-locked loop~\cite{Minder2019, Wang2019, Pittaluga2021, Wang2022}, optical injection locking~\cite{Fang2020, Pittaluga2025, Zheng2026}, and Pound-Drever-Hall technique~\cite{Liu2019, Chen2020, Chen2021twin, Zhou2023, Liu2023}. In these setups, Alice’s and Bob’s laser sources are phase-locked to a reference laser over additional fiber channels, forming a gigantic Mach-Zehnder interferometer. Other resource requirements include ultra-stable cavity, dense wavelength division multiplexing, and proportional-integral-derivative (PID) feedback control. In other experiments, laser phase noise is reconciled without phase locking the two sources~\cite{Li2023, Zhou2023, Chen2024}. For example, Ref.~\cite{Li2023} achieves this over symmetric channels with the use of strong reference pulses at short intervals and a fast Fourier transform algorithm to reconcile the phase difference in postprocessing. In Ref.~\cite{Zhou2023}, Alice and Bob employ sub-Hz linewidth lasers making it possible to cancel the laser frequency difference by using PID feedback control at Charlie's measurement station. In Ref.~\cite{Chen2024}, a local optical frequency standard (the saturated absorption spectroscopy of acetylene) is employed at each source to establish an absolute reference that is sufficient for stable interference over symmetric channels without further beat-note frequency stabilization. 

Similarly, the fiber phase noise is mitigated either in real time~\cite{Minder2019, Wang2019, Chen2020, Pittaluga2021, Wang2022, Zhou2023, Pittaluga2025} or through postprocessing methods~\cite{Liu2019, Fang2020, Liu2021, Chen2021twin, Park2022, Li2023, Liu2023, Chen2024, Zheng2026}. In both cases, a portion of the communication period is reserved for interfering strong reference pulses to measure phase noise. These pulses do not contribute to key generation or quantum bit error rate estimation and this method requires the use of SNSPDs with ultra-short dead times. As the phase noise increases due to longer fiber channels or greater channel asymmetry, so too must the reference pulse detection rate in order to accurately measure the phase difference. 

The aforementioned systems make it difficult to implement a TFQKD network. Firstly, it is practically challenging to stabilize interferometers with highly asymmetric long-fiber links, a task that has only been demonstrated using sub-Hz linewidth lasers with matched frequencies~\cite{Zhou2023}, and this situation is inevitable in a multiuser-pair TFQKD network. Moreover, laser-locking methods greatly increase the complexity and resource requirements, particularly in a network where many sources need to be locked pairwise. Lastly, the SNSPDs required to achieve sufficient count rates for measuring phase noise are an order of magnitude more expensive than SPADs.

Unlike the other systems, TFQKD based on a Sagnac interferometer shows promise for easy networking and low-cost implementation due to common-path stable interference~\cite{Zhong2019, Zhong2021, Zhong2022PRApp}. Since there is only one laser source, no frequency locking is needed. Additionally, since the physical path traveled by Alice's pulses is the same as the physical path traveled by Bob's pulses, $\Delta\phi_{laser}(t)$ is always zero even for conventional laser sources (large $\Delta v(t)$) and highly asymmetric channel lengths and losses between Charlie and each user. The common-path feature also means that $\Delta\phi_{fiber}(t)$ is intrinsically small up to a certain Sagnac loop length~\cite{Mandil2025}, since the fiber phase noise affects the interfering counterpropagating pulses equally. As a result, Sagnac TFQKD is simple, low-cost, and scalable. Users may be added by inserting additional encoding stations at arbitrary points along the same fiber ring. In our previous work~\cite{Mandil2025}, we measured the variance of $\Delta\phi_{fiber}(t)$ in a 125-km Sagnac loop to be 0.025~rad$^2$ and demonstrated that this parameter scales as $L^3$, where $L$ is the Sagnac loop length.

In our TFQKD demonstration, the $1\times1$ optical switch is configured such that pulses are on for 280-$\mu s$ then off for 654-$\mu s$, corresponding to a burst duty cycle of 30\%. These carefully-timed bursts are chosen to ensure that the backscattering has had time to decay, leading to a total noise rate of about $1\times10^{-6}$ per pulse by the time we detect the pulses. When detecting weak signals, the detector dark counts and the backscattering noise (which is roughly 3x the dark count rate) have a more significant impact on the SNR. If we employ burst duty cycles of less than 30\% we may further reduce the QBER. While burst patterning effectively reduces the efficiency of signal generation in our system, we remark that this is not a drawback compared to every other TFQKD system where a portion of the communication period is reserved for measuring phase drift. In fact, our duty cycle for key generation is more favorable than several other TFQKD systems. In future demonstrations, wavelength and time division multiplexing can be used to compensate for the burst off-time.

We may examine the trade-off between visibility and requirements on the light sources. For example, using frequency-locked sources, Ref.~\cite{Pittaluga2025} achieves an average QBER of 5\% (corresponding to a visibility of 90\%~\cite{Clivati2022, Minar2008}) over a 254-km channel using SPADs with a dark count rate of about $1 \times 10^{-6}$ per pulse. Our measurement in Ref.~\cite{Mandil2025} gives a phase-noise-limited visibility of 97\% in a 200-km Sagnac loop. Extrapolating from these results, we expect the phase-noise-limited visibility to be >90\% in a 250-km loop. However, when detecting weak signals and using reasonable burst duty cycles, the visibility in our system becomes SNR-limited due to backscattering noise, which is insignificant in the other scheme. Thus, to achieve similarly high visibilities as phase-locked TFQKD systems, we may improve SNR by mitigating backscattering noise even further or by employing SPADs with higher efficiencies or lower dark count rates (e.g., the ID Qube ULN dark count rate is $< 2\times10^{-7}$ per pulse). Our previous work~\cite{Mandil2025} also helps us to assess how much of the QBER in our current demonstration comes from phase noise versus SNR limitations. Our measurements showed a phase-noise-limited visibility of >98\% (corresponding to a QBER of 1\%) in a 125-km Sagnac loop. Here, we observe an average QBER of 5\% and 6\%, respectively, in the Alice-Danny and Alice-Bob pairs. Thus we attribute 4\% (for Alice-Danny) and 5\% (for Alice-Bob) of the QBER to SNR limitations. 

With regards to security, TFQKD systems that employ phase locking must ensure that no side-channel is introduced by the phase locking mechanism, particularly if Charlie's optical signal is injected into the users' lasers. Even if Charlie's optical signal is first converted into an electrical signal, it is still conceivable that this mechanism may present a side-channel. To achieve a fully secure demonstration in Sagnac TFQKD where users receive and modulate light originating from an untrusted source, additional components should be inserted at each user station to guarantee security~\cite{Gisin2006,Zhao2008,Zhao2010,Xu2015}. As discussed in Refs.~\cite{Zhong2019,Zhong2021,Zhong2022PRApp,Mandil2025}, taps, photodetectors, and bandpass filters are necessary for Alice and Bob to detect and limit strong optical injections from the outside, and to filter out side channels, so as to prevent eavesdroppers from probing the sources. VOAs may be used to attenuate the pulses traveling back to Charlie to single-photon level. Commercially available low-loss components may be chosen to minimize the extra losses introduced to each user station. 

In the future, wavelength division multiplexing may be used to avoid incurring modulator losses from inactive users or to implement concurrent key generation between more than one user pair. Furthermore, we may deploy bidirectional amplifiers between the users to compensate for the fiber loss in order to reduce the intensity of pulses launched into the Sagnac loop (and thus reduce the backscattering noise without reducing the signal generation rate). Another approach may entail the use of low-loss antiresonant hollow-core fibers in the segments between Charlie and the nearest users, since Rayleigh backscattering originating in these segments is the most problematic. These fibers produce backscattering levels more than 40-dB lower than standard SMF~\cite{Belleau2021}. Such strategies will be necessary as the overall loop loss increases, e.g., due to adding more users in the network or replacing ULL fiber with regular SMF fiber. Lastly, we may assess the robustness of our polarization stabilization algorithm over longer durations. We remark that our system infrastructure has applicability to other quantum communication protocols that require phase and polarization stability over long fibers, such as quantum secure direct communication~\cite{Pan2025} and quantum fingerprinting~\cite{Zhong2021QFP}.

\section{Conclusion}\label{sec:conclusion}
In our previous Sagnac TFQKD network demonstration~\cite{Zhong2022PRApp}, short fibers (meters) and a VOA were used in the segments between Charlie and the nearest users. Moreover, the total loop length was limited to 10-km where fast fiber phase noise is negligible and the polarization remains stable without compensation for the full duration of a key generation session. Our current work addresses each of these fiber noise sources to demonstrate a ring TFQKD network that spans 127-km. By implementing a novel scheme for active polarization control, we showed a stable Sagnac interference visibility of $93\pm1\%$ over one hour. Using measured data for the signal basis and simulated data for the decoy basis, we achieved an asymptotic secure key rate of $1.398\times10^{-5}$~bits per pulse over an asymmetric communication channel with 102-km fiber and 45-dB overall loss without any active phase stabilization or postcompensation. Our proof-of-principle demonstration can be made fully secure by inserting additional commercial and low-loss components at each user station. Our experiment represents a highly practical and cost-effective approach to long-distance QKD networks.

\section*{Acknowledgements}
We thank Marcos Curty, Olgierd Zurek, and Yu-Tung Tsai for helpful discussions. We thank Dr. Ming-Jun Li from Corning Inc. for loaning some of the ULL spools. Y.-A.S. acknowledges financial support from Prof. Chih-Sung Chuu. This work is supported by funding from CRC, NSERC, CFI, ORF, and MITACS. Hoi-Kwong Lo is also supported by NUS start-up funding and CQT PI grant. 

\section*{Data availability}
Data underlying the results presented in this paper are not publicly available at this time but may be obtained from the authors upon reasonable request.

\section*{Disclosures}
The authors declare no conflicts of interest.

\appendix

\section{Theory and algorithm of polarization stabilization scheme}\label{app:pol_theory} 
\begin{figure} [htbp]
    \begin{center}
       \includegraphics[width=0.99\linewidth]{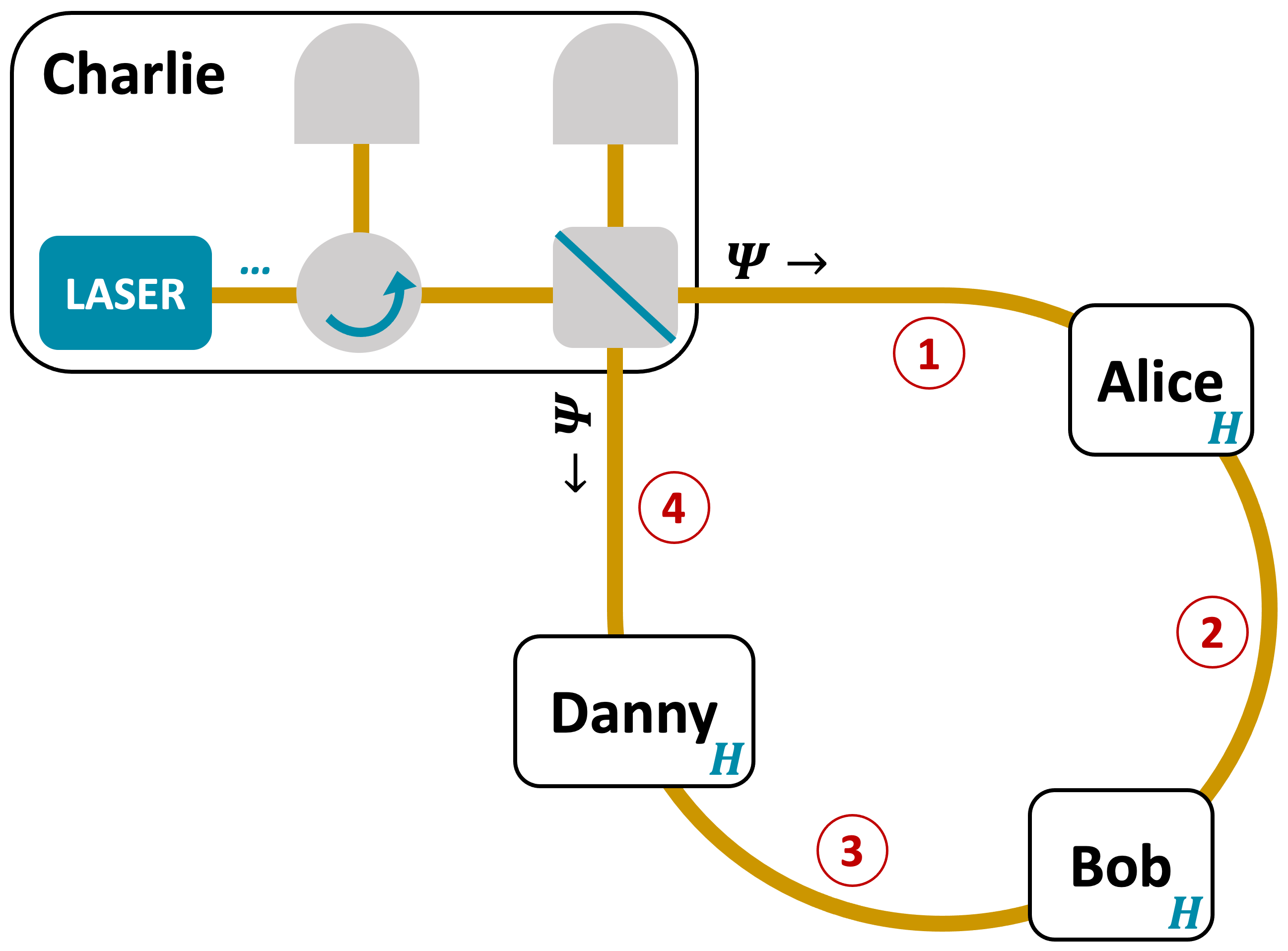} 
    \end{center}
    \caption{Schematic diagram of Sagnac twin-field quantum key distribution network. Each user station acts as a linear polarizer with the polarization denoted $H$. The polarization launched into the Sagnac loop is denoted $\Psi$. The four fiber segments connecting stations are labeled numerically.} \label{fig:pol_scheme}
\end{figure}

In this Appendix, we detail the theory and algorithm behind our polarization stabilization scheme. Fig.~\ref{fig:pol_scheme} depicts a schematic diagram of a three-user-pair Sagnac TFQKD network. All three user stations act as linear polarizers with the polarization denoted $H$. In this section, $H$ and $\Psi$ are two-dimensional complex vectors representing polarization states, and $U$ are $2\times2$ unitary matrices representing the polarization transformation induced by a fiber segment.

Interfering light fields should have identical polarizations in order to maximize the interference visibility. Furthermore, the polarizations of the counterpropagating light fields should be aligned to the linear polarizer inside the user stations. Mathematically, we represent these requirements with the target conditions:
\begin{align}
    U_1^{cc} H = U_4^c H \label{eq:target1} \\
    U_1^c \Psi = H \label{eq:target2} \\
    U_2^c H = H  \label{eq:target3} \\
    U_3^c H = H \label{eq:target4} \\
    U_4^{cc} \Psi = H \label{eq:target5} \\
    U_3^{cc} H = H \label{eq:target6} \\
    U_2^{cc} H = H, \label{eq:target7}
\end{align} where the superscripts $c$ and $cc$ denote the clockwise and counterclockwise path operations, respectively. The subscripts refer to the fiber segments in the Sagnac loop. The polarization launched into the Sagnac loop is denoted $\Psi$. Furthermore, the path operations satisfy the following properties:
\begin{align}
    U U^\dagger = U^\dagger U = I \label{eq:prop1} \\
    U_x^{cc} = (U_x^c)^T, \label{eq:prop2} 
\end{align} where the first condition is unitarity and the second is due to fiber birefringence~\cite{Mecozzi2011}. By~\ref{eq:prop1}, a matrix $U$ that preserves the polarization state $H$ must be a diagonal matrix.

\textbf{\textit{Claim.}} If~\ref{eq:target2} and~\ref{eq:target5} and \underline{one} of~\ref{eq:target3} and~\ref{eq:target7} and \underline{one} of~\ref{eq:target4} and~\ref{eq:target6} are satisfied, then all of~\ref{eq:target1}--~\ref{eq:target7} are satisfied. 

\textbf{\textit{Proof.}} Assume~\ref{eq:target2} and~\ref{eq:target5} are satisfied: $U_1^c \Psi = H$ and $U_4^{cc} \Psi = H$. Then,
\begin{align*}
    \Psi = (U_1^c)^\dagger H \\
    U_4^{cc} (U_1^c)^\dagger H = H \\
    \Rightarrow U_4^{cc} (U_1^c)^\dagger = (U_4^{cc} (U_1^c)^\dagger)^T \quad (\text{symmetric by~\ref{eq:prop1}}) \\
    \Rightarrow (U_4^{cc} (U_1^c)^\dagger)^T = (U_1^{cc})^\dagger U_4^{c} \quad \text{(by~\ref{eq:prop2})} \\
    \Rightarrow (U_1^{cc})^\dagger U_4^{c} H = H \\
    \Rightarrow U_4^{c} H = U_1^{cc} H, 
\end{align*} and~\ref{eq:target1} is satisfied. Assume~\ref{eq:target3} is satisfied: $U_2^c H = H$. Then,
\begin{align*}
    \Rightarrow U_2^c = (U_2^c)^T \quad (\text{symmetric by~\ref{eq:prop1}}) \\
    \Rightarrow (U_2^c)^T = U_2^{cc} \quad \text{(by~\ref{eq:prop2})} \\
    \Rightarrow U_2^{cc} H = H, 
\end{align*} and~\ref{eq:target7} is satisfied. Similarly, if~\ref{eq:target4} is satisfied then~\ref{eq:target6} is satisfied.

In general for a Sagnac TFQKD network with $n$ user stations containing polarizing devices, there are $2n+1$ conditions for polarization alignment and they may be satisfied with $n+1$ operations. That is, by invoking the symmetry between the forward and backward path operations, only one user in the network needs to monitor both directions. For example, the network implemented in this paper relies on four operations (i.e., four intensity monitors and EPCs) for polarization alignment. 

We developed an automated feedback control system that uses a custom-built Arduino-based single photon counter. The feedback loop operates via a modified gradient-ascent optimization algorithm. The system monitors the photon counts at the detectors $D_A$, $D_{B1}$, $D_{B2}$, and $D_D$ every 500-ms. Each EPC and intensity monitor pair operates independently. To minimize unnecessary modulation, the algorithm incorporates a threshold-based stabilization mechanism. Active adjustment of an EPC is triggered only when the detected count rate falls below a predefined target threshold. During the optimization phase, the algorithm tweaks the driving voltage for one axis of the EPC by a fixed step size. If an adjustment yields a higher count rate, the system continues tuning in the same direction; conversely, if the count rate decreases, the step direction is reversed. To avoid being trapped in local optima, if no improvement is observed after consecutive attempts on a single axis, the algorithm sequentially switches to the next axis of the EPC. The precise 12-bit digital control sequences required to drive the EPC are generated and outputted by a National Instruments data acquisition (DAQ) device.

\section{Two-user demonstration}\label{app:twouser} 
\begin{figure} [htbp]
    \begin{center}
       \includegraphics[width=0.99\linewidth]{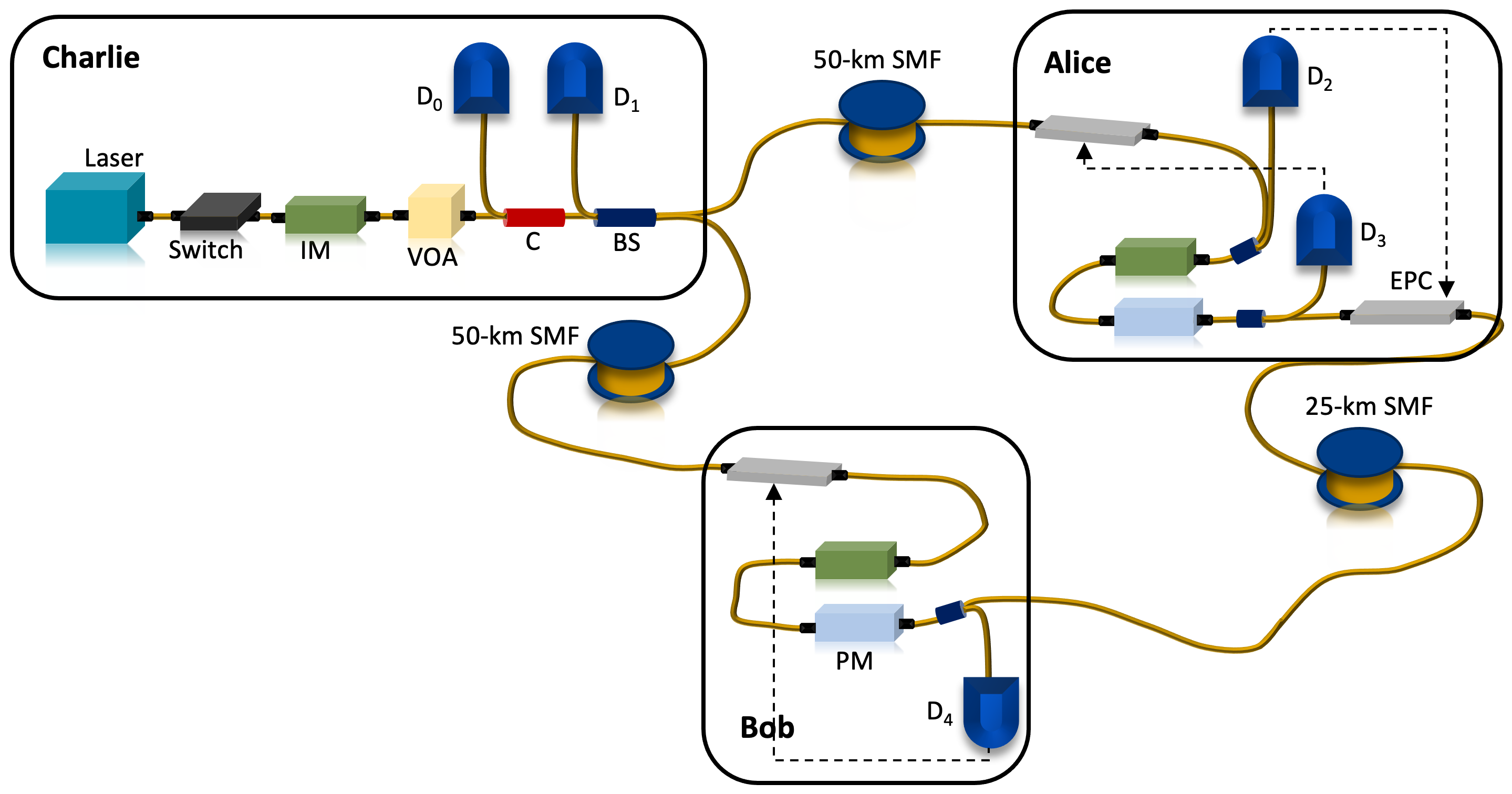} 
    \end{center}
    \caption{Experimental setup of two-user Sagnac twin-field quantum key distribution. IM: intensity modulator; VOA: variable optical attenuator; C: circulator; BS: beam splitter; PM: phase modulator; EPC: electronic polarization controller.} \label{fig:setup_2user}
\end{figure}

In this Appendix, we present the results of a two-user Sagnac TFQKD demonstration where the channels between Charlie and each user are symmetric in both distance and loss. The experimental setup is depicted in Fig.~\ref{fig:setup_2user}. Inside the Sagnac loop, there are two users: Alice and Bob. The total length of the ring is 125-km, made up of three spools of SMF-28 ULL fiber. The procedure is as described in Sec.~\ref{sec:experiment}. Table~\ref{tab:tfqkd_2user} shows the results of our symmetric, two-user TFQKD demonstration over a 100-km channel with 42-dB overall loss. We achieve a QBER (average of $D_0$ and $D_1$) of 2.92\%. The key rates in the infinite-data and finite-data cases are, respectively, $6.249\times10^{-5}$ and $1.111\times10^{-5}$ bits per pulse, approaching the repeaterless bound of $8.002\times10^{-5}$ bits per pulse.

\begin{table*}[htbp]
  \centering
  \caption{Parameters and key rates (bits per pulse) in two-user Sagnac TFQKD. The loss and distance of the channel between Charlie and each user are listed. The detector efficiency (10\%) is attributed as part of the loss between Charlie and each user. Here $s_A,s_B$ are signal intensities (average photon number per pulse) for Alice and Bob, respectively; $\mu,\nu$ are decoy intensities; $P_s,P_\mu,P_\nu$ are probabilities of sending intensities; QBER is quantum bit error rate. The data size for the finite-data case is $10^{10}$.}
  \label{tab:tfqkd_2user}
  \resizebox{\textwidth}{!}{
  \begin{tabular}
  {cccccccccccccc}
  \toprule
  \multicolumn{2}{c}{\textbf{Channel (Alice + Bob)}}&\multicolumn{7}{c}{\textbf{Intensity}}&\multicolumn{2}{c}{\textbf{QBER}}&\multicolumn{3}{c}{\textbf{Key rate} ($\times10^{-5}$ bits per pulse)}\\
  \cmidrule(lr){1-2}\cmidrule(lr){3-9}\cmidrule(lr){10-11}\cmidrule(lr){12-14}
  Loss (dB) & Distance (km) & $s_A$ & $s_B$ & $\mu$ & $\nu$ & $P_s$ & $P_\mu$ & $P_\nu$ & $D_0$ & $D_1$ & Infinite data & Finite data & Repeaterless bound \\
  \midrule
  21+21 & 50+50 & 0.014(2) & 0.014(2) & 0.5120 & 0.1469 & 0.8248 & 0.0174 & 0.0496 & 2.88\% & 2.95\% & 6.249 & 1.111 & 8.002 \\
  \bottomrule
  \end{tabular}
  }
\end{table*}


\bibliography{refs}

\end{document}